# Fundamental principles and applications of nonlinear optical phenomena in classical and quantum electrodynamics

Masud Mansuripur

Wyant College of Optical Sciences, The University of Arizona, Tucson

**Abstract**. Nonlinear optical phenomena play important roles in the vast emerging fields of micro- and nano-technology. This paper describes the general characteristics of nonlinear optical materials and systems, with a focus on parametric amplification, frequency-doubling with pump depletion, quantum noise accompanying attenuation and amplification of light beams, and parametric fluorescence.

**1. Introduction**. Starting in 1960, the advent of lasers with their intense, nearly monochromatic, highly coherent optical radiation opened the door to investigations of nonlinear optical phenomena.[1] While frequency-doubling (also known as second-harmonic generation) has been the prototypical nonlinear optical effect, many other nonlinear phenomena are known and have been well investigated at this point, among them, third harmonic generation, high-harmonic generation, sum- and difference-frequency generation, optical parametric amplification, spontaneous parametric down-conversion, parametric fluorescence, self-focusing, self-phase modulation, four-wave mixing, and optical phase-conjugation.[2-7] Aside from their role in the fundamental studies of classical and quantum optics, nonlinear phenomena form the basis of numerous modern as well as emerging technologies such as ultrafast switches, laser amplifiers, sensors, saturable absorbers, mode-locked lasers, terahertz photonics, supercontinuum light sources, quantum computing, frequency-comb generation, laser material processing, quantum communication, advanced spectroscopy, and ultra-short-pulse lasers.[6-8] The goal of the present paper is to provide an elementary introduction to the subject, aiming to facilitate the journey from the classical to the quantum theory of nonlinear optics.

The next section presents a general overview of the characteristic features of nonlinear optical media, with emphasis on the relation between the material polarization and the local electric field inside the material medium. Then, in Sec.3, we examine the case of multiple electromagnetic (EM) plane-waves of differing frequencies co-propagating inside a transparent nonlinear material. Here, we solve Maxwell's equations for the coupled EM waves and arrive at the governing equations that embody the salient features of nonlinear optical interactions between the fields and the induced electric dipoles. Of particular importance is the notion of phase-matching introduced in this section, followed by a demonstration of energy conservation and the classical Manley-Rowe relations that herald the underlying quantum nature of the nonlinear phenomena under consideration.

Section 4 is devoted to an analysis of parametric amplification, where a weak signal of frequency $\omega_1$ extracts energy from a strong (co-propagating) beam of frequency $\omega_3$, giving rise in the process to a so-called "idler" beam of frequency $\omega_2 = \omega_3 - \omega_1$. Assuming the phase-matching condition is satisfied, both the signal and the idler grow exponentially along the propagation direction. Also discussed in Sec.4 is the case of a nonzero idler entering the nonlinear crystal along with a pump and a signal beam, then exchanging energy (i.e., photons) with them in the ensuing three-wave mixing process. This brings about a pair of classical equations relating the electric fields of the signal and the idler at the exit facet of the nonlinear crystal to the corresponding $E$-fields at its entrance facet, thus providing an occasion to introduce, by analogy, the similarly structured quantum-optical $E$-field operators. Another instance of situations where the evolution of a classical $E$-field inside a nonlinear medium provides a bridge to the corresponding quantum-optical $E$-field operator is the subject of Sec.5, where degenerate parametric amplification is described as a special case of frequency-doubling with pump depletion.

Having motivated the general form and properties of the quantum $E$-field operators in the preceding sections, Sec.6 begins by describing the quantum noise that accompanies an attenuated optical beam, then extends the discussion to generic optical amplification (parametric or otherwise) by introducing an operator that relates the output $E$-field of an amplifier to its input $E$-field. Here, vacuum fluctuations will be seen to be the root cause of extraneous noise that emerges from an optical amplifier, in ways that parallel the contribution of vacuum noise entering through the unused port of a conventional beam-splitter—the obvious difference being that the beam-splitter is an attenuator rather than an amplifier.

Our final example in Sec.7 pertains to the quintessentially quantum phenomenon of parametric fluorescence, where photons of frequency $\omega_3$ from a high-intensity laser beam entering a nonlinear crystal proceed to generate pairs of photons of frequencies $\omega_1$ and $\omega_2$ (with $\omega_1 + \omega_2 = \omega_3$) in a process that is reminiscent of spontaneous photoemission from excited atoms or molecules. The paper closes with a few concluding remarks in Sec.8.

**2. General characteristics of nonlinear media**. A transparent optical medium that responds locally and instantaneously to an applied electric field $\boldsymbol{E}(\boldsymbol{r}, t)$ produces a polarization $\boldsymbol{P}(\boldsymbol{r}, t)$ within the medium that can be characterized by a rank 2 tensor $[\chi_{ij}^{(1)}]$, a rank 3 tensor $[\chi_{ijk}^{(2)}]$, a rank 4 tensor $[\chi_{ijk\ell}^{(3)}]$, etc., as follows:[6,8]

$$P_i = \varepsilon_0 \sum_j \chi_{ij}^{(1)} E_j + \varepsilon_0 \sum \sum_{j,k} \chi_{ijk}^{(2)} E_j E_k + \varepsilon_0 \sum \sum \sum_{j,k,\ell} \chi_{ijk\ell}^{(3)} E_j E_k E_\ell + \cdots. \tag{1}$$

Here, $\varepsilon_0$ is the permittivity of free space (*SI* units: farad/meter), the index $i$ specifies the Cartesian component (i.e., $x, y, z$) of the local polarization vector, and the indices $j, k, \ell$ refer to the Cartesian components of one or more $E$-field vectors that are simultaneously present in the system. Whereas the linear susceptibilities $\chi_{ij}^{(1)}$ are dimensionless, all higher-order (nonlinear) susceptibilities have dimensions that may be readily inferred from their definitions in accordance with Eq.(1); for instance, all $\chi_{ijk}^{(2)}$ have the dimensions of the inverse electric field, or [meter/volt]. Transparency of the medium implies that all the various susceptibilities are real-valued, and the ignorance of material dispersion in Eq.(1) is implicit in the fact that the susceptibilities are taken to be independent of the oscillation frequencies of the participating $E$-fields. While it is fairly straightforward to account for dispersion in the linear susceptibilities $\chi_{ij}^{(1)}$ by allowing them to depend on the frequency $\omega$ of the corresponding $j$, it is rather cumbersome to keep track of dispersion as related to higher-order susceptibilities when $E$-fields of differing frequencies bring about, or emerge from, the nonlinear behavior. This happens for instance, when the induced polarization $P_i$ in Eq.(1) has a contribution from a term such as $\varepsilon_0 \chi_{ijk}^{(2)} E_j(\omega_1) E_k(\omega_2)$, which could give rise to radiation of frequency $\omega_1 + \omega_2$ or $\omega_1 - \omega_2$. It should be noted that the material polarization need not be parallel to the local electric field as, for instance, in anisotropic media the presence of an $E$-field in one direction can induce dielectric polarization in a different direction. Such a phenomenon can be useful in nonlinear optics when differently oriented $E$-fields within a birefringent host medium are associated with different refractive indices, thus making it possible to enforce (or bring about) a desired phase-matching condition under proper arrangements.

For intensities that are accessible via typical lasers, successive terms in the expansion of Eq.(1) usually drop off quickly, and it suffices to keep only the first- and second-order terms, unless $\chi^{(2)}$ (pronounced khi-2) happens to be zero, in which case the third-order term $\chi^{(3)}$ (pronounced khi-3) must be taken into account. The symmetries of nonlinear media frequently reveal at the outset that many components of the susceptibility tensors must vanish. An important example is provided by centrally-symmetric media which are invariant under inversion through the origin of the coordinates. Inversion with respect to the origin flips the signs of the $E$-field components, but also of the material polarization, which is incompatible with the existence of nonlinear susceptibilities of even order. Thus, in a centrally-symmetric medium, the even-order terms in Eq.(1) vanish, leaving the third-order term as the leading nonlinear contributor to the host material's polarization $P(r,t)$.

**3. Nonlinear optics**. Let a number of plane-waves of differing frequencies $\omega_1, \omega_2, \cdots$, all linearly-polarized along $\hat{x}$, propagate in the $z$-direction within a homogeneous, isotropic and transparent medium of refractive index $n(\omega)$. Each plane-wave will have a $k$-vector $\boldsymbol{k}_\ell = (n_\ell \omega_\ell/c)\hat{\boldsymbol{z}}$, where $n_\ell = n(\omega_\ell)$. Thus, the sum of the $E$-fields of the co-propagating plane-waves may be written as[9,10]

$$\boldsymbol{E}(\boldsymbol{r},t) = \sum_\ell E_{0x,\ell}\hat{\boldsymbol{x}}\, e^{i[(n_\ell \omega_\ell/c)z - \omega_\ell t]}. \tag{2}$$

When two collimated beams, linearly-polarized along $\hat{x}$ and having frequencies $\omega_1$ and $\omega_2$, propagate in the $z$ direction, their total intensity as a function of position and time $(z,t)$ will be

$$I(z,t) = \tfrac{1}{2}(\boldsymbol{E} + \boldsymbol{E}^*) \cdot \tfrac{1}{2}(\boldsymbol{E} + \boldsymbol{E}^*) = \tfrac{1}{4}\boldsymbol{E}\cdot\boldsymbol{E} + \tfrac{1}{4}\boldsymbol{E}^*\cdot\boldsymbol{E}^* + \tfrac{1}{2}\boldsymbol{E}\cdot\boldsymbol{E}^*$$

$$= \tfrac{1}{4}(\boldsymbol{E}_1\cdot\boldsymbol{E}_1 + \boldsymbol{E}_1^*\cdot\boldsymbol{E}_1^*) + \tfrac{1}{4}(\boldsymbol{E}_2\cdot\boldsymbol{E}_2 + \boldsymbol{E}_2^*\cdot\boldsymbol{E}_2^*) + \tfrac{1}{2}(\boldsymbol{E}_1\cdot\boldsymbol{E}_2 + \boldsymbol{E}_1^*\cdot\boldsymbol{E}_2^*)$$

$$+ \tfrac{1}{2}(\boldsymbol{E}_1\cdot\boldsymbol{E}_1^* + \boldsymbol{E}_2\cdot\boldsymbol{E}_2^* + \boldsymbol{E}_1\cdot\boldsymbol{E}_2^* + \boldsymbol{E}_2\cdot\boldsymbol{E}_1^*)$$

$$= \tfrac{1}{2}\mathrm{Re}\{E_{0x,1}^2 e^{i[(2n_1\omega_1/c)z - 2\omega_1 t]}\} \quad\text{[frequency doubling: } \omega_3 = 2\omega_1\text{]} + \tfrac{1}{2}\mathrm{Re}\{E_{0x,2}^2 e^{i[(2n_2\omega_2/c)z - 2\omega_2 t]}\} \quad\text{[frequency doubling: } \omega_3 = 2\omega_2\text{]}$$

$$+ \mathrm{Re}\{E_{0x,1}E_{0x,2}\, e^{i[(n_1\omega_1+n_2\omega_2)(z/c) - (\omega_1+\omega_2)t]}\} \leftarrow \text{frequency addition: } \omega_3 = \omega_1 + \omega_2$$

$$+ \mathrm{Re}\{E_{0x,1}E_{0x,2}^*\, e^{i[(n_1\omega_1-n_2\omega_2)(z/c) - (\omega_1-\omega_2)t]}\} \leftarrow \text{frequency subtraction: } \omega_3 = \omega_1 - \omega_2$$

$$+ \tfrac{1}{2}|E_{0x,1}|^2 + \tfrac{1}{2}|E_{0x,2}|^2. \leftarrow \text{optical rectification} \tag{3}$$

The host material responds to the traveling electromagnetic (EM) wave by developing a polarization $P(r,t)$ that, to a first approximation, is proportional to the local electric field $E(r,t)$, with a proportionality constant $\varepsilon_0\chi^{(1)}$ commonly known as the material medium's linear susceptibility.[†] If the traveling wave happens to be intense, however, there will be additional contributions to $P(r,t)$ rooted in the nonlinear response of the host material to the propagating EM field. In the case of non-centrosymmetric material media, for instance, the nonlinear contribution to $P(r,t)$ in the next order of approximation will be proportional to the local $E$-field intensity, with a proportionality constant $\varepsilon_0\chi^{(2)}$, known as the khi-2 susceptibility. Thus, with the $E$-field distribution given by Eq.(2), and with the combined intensity of two of the plane-waves being strong enough to induce sum-frequency generation in accordance with Eq.(3), the relevant material polarization may be expressed as follows:

$$\boldsymbol{P}(\boldsymbol{r},t) = \varepsilon_0\chi^{(1)}\boldsymbol{E}(\boldsymbol{r},t) + \varepsilon_0\chi^{(2)}E_{0x,1}E_{0x,2}\, e^{i[(n_1\omega_1+n_2\omega_2)(z/c) - (\omega_1+\omega_2)t]}\hat{\boldsymbol{x}}. \tag{4}^{\ddagger}$$

---

[†] To account for the presence of dispersion in the host medium, one can take $\chi^{(1)}$ to be a function of the frequency $\omega$, then use $\varepsilon_0\chi^{(1)}(\omega_\ell)$ as the coefficient for each plane-wave that appears in Eq.(2). This will yield a more accurate representation of the linear polarization term in Eq.(4), which, in its current form, is ignorant of dispersion.

[‡] Unlike $\chi^{(1)}$, which is dimensionless, $\chi^{(2)}$ has the dimensions of the inverse of the $E$-field, namely, meter/volt in *SI*.

The presence of a term with the frequency $\omega_3 = \omega_1 + \omega_2$ in the above expression of the host medium's polarization produces an EM wave of frequency $\omega_3$, which must be included in the $E$-field profile of Eq.(2). Considering that the strengths of the various waves appearing in Eq.(2) may now vary with the propagation distance $z$ (in consequence of their nonlinear interactions), we allow the amplitudes $E_{0x,\ell}$ to be functions of $z$, and proceed to estimate $E_{0x,\ell}(z)$ by solving the relevant Maxwell's equations.[9,10] From $\boldsymbol{\nabla} \times \boldsymbol{H} = \partial \boldsymbol{D}/\partial t$ and $\boldsymbol{\nabla} \times \boldsymbol{E} = -\partial \boldsymbol{B}/\partial t$, given that $\boldsymbol{B} = \mu_0 \boldsymbol{H}$ and $\boldsymbol{\nabla} \times (\boldsymbol{\nabla} \times \boldsymbol{E}) = \boldsymbol{\nabla}(\boldsymbol{\nabla} \cdot \boldsymbol{E}) - \boldsymbol{\nabla}^2 \boldsymbol{E}$, and also that $\boldsymbol{\nabla} \cdot \boldsymbol{E} = 0$,[§] we arrive at the following wave equation:

$$\boldsymbol{\nabla}^2 \boldsymbol{E}(\boldsymbol{r},t) = \mu_0 \frac{\partial^2}{\partial t^2} \boldsymbol{D}(\boldsymbol{r},t). \tag{5}$$

Substitution from Eq.(2) into the left-hand side of Eq.(5) yields

$$\boldsymbol{\nabla}^2 \boldsymbol{E}(\boldsymbol{r},t) = \sum_\ell \left[ \frac{\mathrm{d}^2 E_{0x,\ell}(z)}{\mathrm{d}z^2} + \mathrm{i}(2n_\ell \omega_\ell/c) \frac{\mathrm{d}E_{0x,\ell}(z)}{\mathrm{d}z} - (n_\ell \omega_\ell/c)^2 E_{0x,\ell}(z) \right] e^{\mathrm{i}[(n_\ell \omega_\ell/c)z - \omega_\ell t]} \hat{\boldsymbol{x}}. \tag{6}$$

In what follows, we shall assume that $E_{0x,\ell}$ is a slowly-varying function of $z$, whose second derivative with respect to $z$ is small and can, therefore, be dropped from Eq.(6). As for the right-hand side of Eq.(5), we invoke the identities $\boldsymbol{D} = \varepsilon_0 \boldsymbol{E} + \boldsymbol{P}$ and $n^2(\omega) = 1 + \chi^{(1)}(\omega)$ in conjunction with Eq.(4) to write

$$\boldsymbol{D}(\boldsymbol{r},t) = \varepsilon_0 \{ \sum_\ell n_\ell^2 E_{0x,\ell} e^{\mathrm{i}[(n_\ell \omega_\ell/c)z - \omega_\ell t]} + \chi^{(2)} E_{0x,1} E_{0x,2} e^{\mathrm{i}[(n_1 \omega_1 + n_2 \omega_2)(z/c) - (\omega_1 + \omega_2)t]} \} \hat{\boldsymbol{x}}. \tag{7}$$

Suppose now that the (large) amplitudes $E_{0x,1}$ and $E_{0x,2}$ of the pump beams remain more or less constant along the propagation direction. The only beam whose amplitude's $z$-dependence must be accounted for is thus the relatively weak beam having the sum frequency $\omega_3 = \omega_1 + \omega_2$ and the amplitude $E_{0x,3}(z)$. Substituting from Eqs.(6) and (7) into Eq.(5), ignoring the second derivative of $E_{0x,3}(z)$, and recalling that the speed of light in vacuum is given by $c = (\mu_0 \varepsilon_0)^{-\frac{1}{2}}$, we find

$$\mathrm{i}\left(\frac{2n_3 \omega_3}{c}\right) \frac{\mathrm{d}E_{0x,3}(z)}{\mathrm{d}z} e^{\mathrm{i}[(n_3 \omega_3/c)z - \omega_3 t]} = -\frac{\chi^{(2)}}{c^2}(\omega_1 + \omega_2)^2 E_{0x,1} E_{0x,2} e^{\mathrm{i}[(n_1 \omega_1 + n_2 \omega_2)(z/c) - (\omega_1 + \omega_2)t]}. \tag{8}$$

Further simplification of the above equation yields

$$\frac{\mathrm{d}E_{0x,3}(z)}{\mathrm{d}z} = \frac{\mathrm{i}\chi^{(2)} \omega_3 E_{0x,1} E_{0x,2}}{2n_3 c} e^{\mathrm{i}(n_1 \omega_1 + n_2 \omega_2 - n_3 \omega_3)(z/c)}. \tag{9}$$

If we now set to zero the amplitude $E_{0x,3}$ of the sum-frequency wave at the entrance facet of the nonlinear crystal (i.e., at $z = 0$), we obtain the solution of Eq.(9), as follows:

$$E_{0x,3}(z) = \frac{\chi^{(2)} \omega_3 E_{0x,1} E_{0x,2}}{2n_3 c} \left( \frac{e^{\mathrm{i}(k_1 + k_2 - k_3)z} - 1}{k_1 + k_2 - k_3} \right)$$

$$= \mathrm{i} e^{\mathrm{i}(k_1 + k_2 - k_3)z/2} \left( \frac{\chi^{(2)} \omega_3 E_{0x,1} E_{0x,2}}{n_3 c} \right) \left( \frac{\sin[(k_1 + k_2 - k_3)z/2]}{k_1 + k_2 - k_3} \right). \tag{10}$$

The above solution reaches its peak value at $z = \pi/|k_1 + k_2 - k_3|$, which is an ideal choice for the thickness of the nonlinear crystal. However, recalling that $\sin x \cong x$ for sufficiently small values

---

[§] In the absence of free electric charge-density $\rho_{\text{free}}$, the Maxwell equation $\boldsymbol{\nabla} \cdot \boldsymbol{D} = \rho_{\text{free}}$ becomes $\boldsymbol{\nabla} \cdot (\varepsilon_0 \boldsymbol{E} + \boldsymbol{P}) = 0$. In isotropic, linear, and homogeneous media, where $\boldsymbol{P}(\boldsymbol{r})e^{-\mathrm{i}\omega t} = \varepsilon_0 \chi^{(1)}(\omega)\boldsymbol{E}(\boldsymbol{r})e^{-\mathrm{i}\omega t}$, one concludes that $\boldsymbol{\nabla} \cdot \boldsymbol{E}(\boldsymbol{r}) = 0$. However, when the host medium is anisotropic or nonlinear or inhomogeneous, the divergence of the $E$-field can no longer be assumed to vanish. Nevertheless, when the sole $E$-field components $(E_x, E_y)$ are confined to the $xy$-plane and vary only with $(z,t)$, as is the case in the present paper, it is allowed to set $\boldsymbol{\nabla} \cdot \boldsymbol{E}(\boldsymbol{r},t) = (\partial E_x/\partial x) + (\partial E_y/\partial y) = 0$.

of $x$, if one manages to satisfy the phase-matching condition, namely, $k_3 = k_1 + k_2$, then the sum-frequency signal will grow linearly with the propagation distance $z$. (Keep in mind that the approximation leading to Eq.(10) will eventually break down when the pump amplitudes $E_{0x,1}$ and $E_{0x,2}$ can no longer be assumed to remain constant along the $z$-axis.)

Let us now examine the coupled variations of the pump and signal waves by bringing the $z$-dependences of $E_{0x,1}, E_{0x,2}$ under consideration. The intensity of the combined $\boldsymbol{E}_1 + \boldsymbol{E}_3$ produces a signal with the difference-frequency $\omega_2 = \omega_3 - \omega_1$ in accordance with an equation akin to Eq.(3). Since the phase-matching condition $k_2 = k_3 - k_1$ for this wave is the same as that for the previously considered sum-frequency signal, it too can propagate unhindered along the $z$-axis. Similarly, the intensity of the combined $\boldsymbol{E}_2 + \boldsymbol{E}_3$ gives rise to a signal with the difference-frequency $\omega_1 = \omega_3 - \omega_2$, which is also phase-matched and can, therefore, readily propagate within our khi-2 crystal. We assume that the remaining sum and difference frequencies (such as $\omega_1 + \omega_3$ or $\omega_1 - \omega_2$) as well as all doubled frequencies (e.g., $2\omega_1$ or $2\omega_2$) are not properly phase-matched and that, therefore, they fail to materialize within the host medium. All in all, with the nonlinear polarization terms corresponding to the difference frequencies $\omega_2 = \omega_3 - \omega_1$ and $\omega_1 = \omega_3 - \omega_2$ properly incorporated into Eq.(7), we obtain from Eq.(5), in conjunction with Eqs.(6) and (7), the following coupled differential equations for $E_{0x,1}$, $E_{0x,2}$, and $E_{0x,3}$:[8]

$$\frac{dE_{0x,1}(z)}{dz} = i \frac{\chi^{(2)} \omega_1}{2 n_1 c} E_{0x,3}(z) E^*_{0x,2}(z) e^{-i(k_1+k_2-k_3)z}, \tag{11}$$

$$\frac{dE_{0x,2}(z)}{dz} = i \frac{\chi^{(2)} \omega_2}{2 n_2 c} E_{0x,3}(z) E^*_{0x,1}(z) e^{-i(k_1+k_2-k_3)z}, \tag{12}$$

$$\frac{dE_{0x,3}(z)}{dz} = i \frac{\chi^{(2)} \omega_3}{2 n_3 c} E_{0x,1}(z) E_{0x,2}(z) e^{i(k_1+k_2-k_3)z}. \tag{13}$$

Needless to say, when the phase-matching is exact, the phase-factors $e^{\pm i(k_1+k_2-k_3)z}$ drop out of the above equations.

For each wave inside the crystal, the time-averaged Poynting vector $\boldsymbol{S}_\ell = n_\ell |E_{0x,\ell}|^2 \hat{\boldsymbol{z}}/(2 Z_0)$ [$Z_0 = (\mu_0/\varepsilon_0)^{½}$ is the impedance of free space] corresponding to the rate of flow of optical energy (per unit area per unit time) can be differentiated with respect to $z$, yielding

$$\frac{dS_{z,\ell}}{dz} = \frac{n_\ell}{2 Z_0} \left( \frac{dE_{0x,\ell}}{dz} E^*_{0x,\ell} + E_{0x,\ell} \frac{dE^*_{0x,\ell}}{dz} \right). \tag{14}$$

Substitution for $dE_{0x,\ell}/dz$ and its complex-conjugate from Eqs.(11)-(13) into Eq.(14) yields

$$\frac{dS_{z,1}}{dz} = ½ \varepsilon_0 \chi^{(2)} \omega_1 \text{Im}\left[ E_{0x,1} E_{0x,2} E^*_{0x,3} e^{i(k_1+k_2-k_3)z} \right]. \tag{15}$$

$$\frac{dS_{z,2}}{dz} = ½ \varepsilon_0 \chi^{(2)} \omega_2 \text{Im}\left[ E_{0x,1} E_{0x,2} E^*_{0x,3} e^{i(k_1+k_2-k_3)z} \right]. \tag{16}$$

$$\frac{dS_{z,3}}{dz} = -½ \varepsilon_0 \chi^{(2)} \omega_3 \text{Im}\left[ E_{0x,1} E_{0x,2} E^*_{0x,3} e^{i(k_1+k_2-k_3)z} \right]. \tag{17}$$

Adding up the above equations now reveals that the overall energy of the system is conserved; that is,

$$\frac{d}{dz}(S_{z,1} + S_{z,2} + S_{z,3}) = ½(\omega_1 + \omega_2 - \omega_3) \varepsilon_0 \chi^{(2)} \text{Im}\left[ E_{0x,1} E_{0x,2} E^*_{0x,3} e^{i(k_1+k_2-k_3)z} \right] = 0. \tag{18}$$

To state the content of Eq.(18) in words, any reduction in the energy flux of the pump beams is directly accounted for by the increase in the energy flux of the sum-frequency signal.

Another important conclusion that one may draw from Eqs.(15)-(17) is the so-called Manley-Rowe relations,[8] which are expressed as follows:

$$\frac{1}{\omega_1}\frac{dS_{z,1}}{dz} = \frac{1}{\omega_2}\frac{dS_{z,2}}{dz} = -\frac{1}{\omega_3}\frac{dS_{z,3}}{dz}. \tag{19}$$

The straightforward interpretation of the Manley-Rowe relations is that, within any interval $\Delta z$ along the $z$-axis and during any time window $\Delta t$, the number of photons extracted from the first pump beam equals the number of photons extracted from the second pump beam, which in turn equals the number of photons acquired by the third (signal) beam.

**4. Parametric amplification**. Suppose an intense beam of frequency $\omega_3$ (the pump) is launched into a nonlinear khi-2 medium along with a weak beam of frequency $\omega_1$ (the signal). Assuming the phase-matching condition for difference-frequency generation is satisfied, a beam of frequency $\omega_2 = \omega_3 - \omega_1$ (commonly referred to as the "idler") is produced that propagates in parallel with the pump and the signal along the $z$-direction; see Fig.1. The phase-matching condition is $k_2 = k_3 - k_1$, where, as before, $\boldsymbol{k}_\ell = (n_\ell \omega_\ell/c)\hat{\boldsymbol{z}}$. Thus, at the entrance facet of the host medium (i.e., at $z = 0$), the initial conditions are specified as some fixed, albeit small, value $E_{0x,1}(0)$ for the signal, $E_{0x,2}(0) = 0$ for the idler, and another fixed, albeit large, value $E_{0x,3}(0)$ for the pump. Note that the way we have enumerated the beams in this three-wave mixing situation coincides with the enumeration used in the preceding section, where $\omega_3$ was the sum of the frequencies $\omega_1$ and $\omega_2$. Consequently, we may continue to use the coupled equations (11)-(13) to address the present example of parametric amplification.

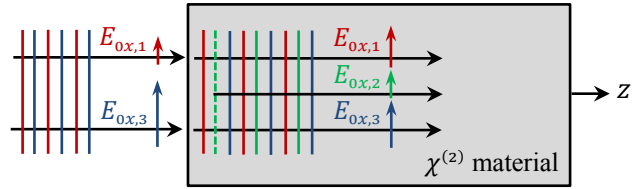

**Fig.1**. Parametric amplification of a signal wave of frequency $\omega_1$ and amplitude $E_{0x,1}$ via nonlinear interaction (within a $\chi^{(2)}$ medium) with a pump beam of frequency $\omega_3$ and amplitude $E_{0x,3}$. In the process, an idler wave with the difference frequency $\omega_2 = \omega_3 - \omega_1$ and co-propagating with the pump and signal in the $z$-direction is generated—provided, of course, that the phase-matching condition $k_2 = k_3 - k_1$ is satisfied. Both the signal and the idler grow in magnitude at the expense of the pump beam. The classical Manley-Rowe relations anticipate the quantum interpretation that each pump photon of energy $\hbar\omega_3$ splits into a pair of photons of energies $\hbar\omega_1$ and $\hbar\omega_2$ that feed the growing signal and idler beams.

Assuming that the pump beam remains more or less constant along the $z$-axis (i.e., the signal and the idler are too weak to extract much energy from the pump), we may ignore Eq.(13), treat $E_{0x,3}(z)$ as a constant, drop the phase-factor $e^{-\mathrm{i}(k_1+k_2-k_3)z}$ from Eqs.(11) and (12) (i.e., take the phase-matching to be perfect), and proceed to combine these equations to arrive at

$$\frac{d^2 E_{0x,1}(z)}{dz^2} = \zeta^2 E_{0x,1}(z), \quad \text{where} \quad \zeta = \frac{\chi^{(2)}|E_{0x,3}|}{2c}\left(\frac{\omega_1 \omega_2}{n_1 n_2}\right)^{1/2}. \tag{20}$$

We find from Eq.(11) that $dE_{0x,1}/dz$ is zero at $z = 0$. Consequently, the solution of Eq.(20) is

$$E_{0x,1}(z) = E_{0x,1}(0)\cosh(\zeta z). \tag{21}$$

Substituting from Eq.(21) into Eq.(12) and recalling that $E_{0x,2}(0) = 0$, we find

$$E_{0x,2}(z) = \frac{iE_{0x,3}E_{0x,1}^*(0)}{|E_{0x,3}|}\left(\frac{n_1\omega_2}{n_2\omega_1}\right)^{½}\sinh(\zeta z). \tag{22}$$

It is thus seen that the signal amplitude $E_{0x,1}$ grows with the distance $z$ away from the entrance facet, while the idler, which is non-existent at first (i.e., at $z = 0$), similarly grows along the $z$-axis. Eventually, of course, the signal and the idler become large enough to begin to impact the strength of the pump, by which point the approximation of constant $E_{0x,3}$ breaks down.

Parametric amplification is a useful process since it can be tuned (by adjusting the phase-matching condition) over a broad range of both frequencies $\omega_1$ and $\omega_3$. Optical parametric amplification (OPA) has been used to produce tunable lasers (known as optical parametric oscillators or OPOs) by placing a nonlinear host medium inside a resonant Fabry-Pérot cavity.[6-8]

If the initial condition for the idler at $z = 0$ happens to be $E_{0x,2}(0) \neq 0$, Eqs.(11) and (12), again under the assumption that the pump is constant, lead to the following solutions for the signal and the idler:

$$E_{0x,1}(z) = E_{0x,1}(0)\cosh(\zeta z) + i\left(\frac{n_2\omega_1}{n_1\omega_2}\right)^{½}\frac{E_{0x,3}}{|E_{0x,3}|}E_{0x,2}^*(0)\sinh(\zeta z), \tag{23a}$$

$$E_{0x,2}(z) = E_{0x,2}(0)\cosh(\zeta z) + i\left(\frac{n_1\omega_2}{n_2\omega_1}\right)^{½}\frac{E_{0x,3}}{|E_{0x,3}|}E_{0x,1}^*(0)\sinh(\zeta z). \tag{23b}$$

These equations are somewhat simplified if we write $E_{0x,3}/|E_{0x,3}| = e^{i\varphi_3}$, then express the emergent signal and idler $E$-fields at the output of a parametric amplifier of length $L$ as follows:

$$\boldsymbol{E}_1(x,y,z=L,t) = \left[E_{0x,1}(0)\cosh(\zeta L) + i\left(\frac{n_2\omega_1}{n_1\omega_2}\right)^{½}e^{i\varphi_3}E_{0x,2}^*(0)\sinh(\zeta L)\right]\hat{x}e^{i[n_1\omega_1(L/c)-\omega_1 t]}, \tag{24a}$$

$$\boldsymbol{E}_2(x,y,z=L,t) = \left[E_{0x,2}(0)\cosh(\zeta L) + i\left(\frac{n_1\omega_2}{n_2\omega_1}\right)^{½}e^{i\varphi_3}E_{0x,1}^*(0)\sinh(\zeta L)\right]\hat{x}e^{i[n_2\omega_2(L/c)-\omega_2 t]}. \tag{24b}$$

These purely classical EM expressions have the correct structure for being converted to the corresponding quantum $E$-field operators[6,8,11] at the exit facet of the amplifier, namely,

$$\hat{\boldsymbol{E}}_1^{(+)}(L) = e^{in_1\omega_1 L/c}\left[\cosh(\zeta L)\hat{E}_{x,1}^{(+)}(0) + i\left(\frac{n_2\omega_1}{n_1\omega_2}\right)^{½}e^{i\varphi_3}\sinh(\zeta L)\hat{E}_{x,2}^{(-)}(0)\right]\hat{x}, \tag{25a}$$

$$\hat{\boldsymbol{E}}_2^{(+)}(L) = e^{in_2\omega_2 L/c}\left[\cosh(\zeta L)\hat{E}_{x,2}^{(+)}(0) + i\left(\frac{n_1\omega_2}{n_2\omega_1}\right)^{½}e^{i\varphi_3}\sinh(\zeta L)\hat{E}_{x,1}^{(-)}(0)\right]\hat{x}. \tag{25b}$$

One can verify that the above single-mode $E$-field operators at the amplifier's output obey the correct (i.e., canonical) commutation relations—given a pair of input beams each in the same mode as the corresponding output beam.[8] In particular, recalling that $\cosh^2(\zeta L) - \sinh^2(\zeta L) = 1$, we find

$$\left[\hat{E}_{x,1}^{(+)}(L), \hat{E}_{x,1}^{(-)}(L)\right] = \left[\hat{E}_{x,1}^{(+)}(0), \hat{E}_{x,1}^{(-)}(0)\right], \tag{26a}$$

$$\left[\hat{E}_{x,2}^{(+)}(L), \hat{E}_{x,2}^{(-)}(L)\right] = \left[\hat{E}_{x,2}^{(+)}(0), \hat{E}_{x,2}^{(-)}(0)\right]. \tag{26b}$$

The above derivation has relied on the fact that the commutator of the $E$-field operators for the single-mode plane-waves $(\omega_k, \boldsymbol{k}_k, \hat{\boldsymbol{e}}_k)$ and $(\omega_\ell, \boldsymbol{k}_\ell, \hat{\boldsymbol{e}}_\ell)$ propagating in a medium of refractive index $n(\omega)$ is given by

$$[\hat{\boldsymbol{E}}_k^{(+)}, \hat{\boldsymbol{E}}_\ell^{(-)}] = \left[\left(\frac{\hbar\omega_k}{2\varepsilon_0 n_k V}\right)^{1/2} e^{\mathrm{i}(\boldsymbol{k}_k \cdot \boldsymbol{r} - \omega_k t)} \hat{\boldsymbol{e}}_k \hat{a}_k, \left(\frac{\hbar\omega_\ell}{2\varepsilon_0 n_\ell V}\right)^{1/2} e^{-\mathrm{i}(\boldsymbol{k}_\ell \cdot \boldsymbol{r} - \omega_\ell t)} \hat{\boldsymbol{e}}_\ell^* \hat{a}_\ell^\dagger \right] = \left(\frac{\hbar\omega_k}{2\varepsilon_0 n_k V}\right) \delta_{k,\ell}. \tag{27}$$

(Kronecker's delta-function)

The reason for the refractive index $n(\omega)$ appearing in the denominator of the single-photon $E$-field amplitude is that, in classical electrodynamics, the $\boldsymbol{E}$ and $\boldsymbol{B}$ fields of a propagating plane-wave within a homogeneous medium of refractive index $n$ are related via $\boldsymbol{B} = n\hat{\boldsymbol{k}} \times \boldsymbol{E}/c$, where $\hat{\boldsymbol{k}} = \boldsymbol{k}/k$. Consequently, the asserted single-photon $E$-field amplitude $[\hbar\omega/(2\varepsilon_0 n V)]^{1/2}$ leads to the quantized time-averaged Poynting vector $\langle \boldsymbol{E} \times \boldsymbol{B}/\mu_0 \rangle = (\hbar\omega c/V)\hat{\boldsymbol{k}}$, an entity that is independent of $n$. Note that $V$ continues to represent the volume occupied by the plane-wave in free space (i.e., outside the host medium of refractive index $n$), whereas, upon entering the host medium, the free-space $E$-field must be scaled by $\sqrt{n}$ in order to retain the wave's energy $\hbar\omega$.

**5. Frequency doubling with pump depletion**. Let an intense beam of frequency $\omega_1$ and initial amplitude $E_{0x,1}(0)$ propagate in a nonlinear khi-2 medium along the $z$-axis, producing a second-harmonic signal of frequency $\omega_2 = 2\omega_1$ and amplitude $E_{0x,2}(z)$, the initial condition being $E_{0x,2}(0) = 0$. The intensity of the combined $\boldsymbol{E}_1 + \boldsymbol{E}_2$ field is given by Eq.(3), showing the presence of a term with the doubled-frequency $2\omega_1$ and amplitude $\tfrac{1}{2}E_{0x,1}^2$, in addition to another term with the difference frequency $\omega_2 - \omega_1$ and amplitude $E_{0x,1}^* E_{0x,2}$. (The contributions of the remaining terms with frequencies $2\omega_2$ and $\omega_1 + \omega_2$ are negligible for lack of phase-matching.) The phase-matching condition $k_2 = 2k_1$ for second-harmonic generation leads to $n_2\omega_2/c = 2n_1\omega_1/c$ and, consequently, to $n(\omega_1) = n(\omega_2)$. Equation (5) in conjunction with Eqs.(6) and (7) — again, ignoring the terms $\mathrm{d}^2 E_{0x,\ell}/\mathrm{d}z^2$ and bringing into Eq.(7) the relevant nonlinear factors from Eq.(3) — now yields the coupled evolution equations for $E_{0x,1}$ and $E_{0x,2}$, as follows:

$$\frac{\mathrm{d}E_{0x,1}(z)}{\mathrm{d}z} = \mathrm{i}\frac{\omega_1 \chi^{(2)}}{2n_1 c} E_{0x,1}^* E_{0x,2} e^{\mathrm{i}(k_2 - 2k_1)z}, \tag{28}$$

$$\frac{\mathrm{d}E_{0x,2}(z)}{\mathrm{d}z} = \mathrm{i}\frac{\omega_2 \chi^{(2)}}{4n_2 c} E_{0x,1}^2 e^{-\mathrm{i}(k_2 - 2k_1)z}. \tag{29}$$

The above equations are further simplified in the presence of perfect phase-matching, where the phase-factors $e^{\pm \mathrm{i}(k_2 - 2k_1)z}$ drop out and the leading coefficients on the right-hand sides of the two equations become identical — simply because $n_2 = n_1$ and $\omega_2 = 2\omega_1$. Defining the auxiliary parameter $\zeta = \omega_1 \chi^{(2)} |E_{0x,1}(0)|/(2n_1 c)$, the exact solution of these coupled equations is found to be

$$E_{0x,1}(z) = \frac{E_{0x,1}(0)}{\cosh(\zeta z)}, \tag{30}$$

$$E_{0x,2}(z) = \mathrm{i}\frac{E_{0x,1}^2(0)}{|E_{0x,1}(0)|} \tanh(\zeta z). \tag{31}$$

---

** The significance of this commutation relation is rooted in the fact that $[\hat{a}_\ell, \hat{a}_\ell^\dagger]$ acting on any number state $|n\rangle$ leaves the state intact. Considering that any single-mode beam in an arbitrary pure state $|\psi\rangle$ is in a superposition of various number states, the action of $[\hat{a}_\ell, \hat{a}_\ell^\dagger]$ on $|\psi\rangle$ is not expected to change the state. By the same token, the action of $[\hat{\boldsymbol{E}}_\ell^{(+)}, \hat{\boldsymbol{E}}_\ell^{(-)}]$ on $|\psi\rangle$ should only multiply the state by the single-photon $E$-field intensity $\hbar\omega_\ell/(2\varepsilon_0 n_\ell V)$.

Note that, with an increasing $z$, the pump amplitude drops (slowly at first and then rapidly) toward zero, while the second-harmonic signal rises initially and then saturates. The parameter $\zeta$ controls the rate of transfer of energy from the pump to the second harmonic signal. The sum of the Poynting vectors of the two beams is constant along the entire propagation path.

**Digression**. Equations (28) and (29) may also be used to examine the case of *degenerate* parametric amplification, where an intense pump beam of frequency $\omega_2$ co-propagates with a weaker beam of frequency $\omega_1 = \tfrac{1}{2}\omega_2$ along the $z$-axis. There is, of course, no separate idler wave in this case as the frequencies of the signal ($\omega_1$) and the idler ($\omega_2 - \omega_1$) are identical. (Note that the index 2 is assigned here to the pump, whereas, in the preceding section, 2 was the index of the idler, and 3 served as the index for the pump.) Assuming (as we did in the preceding section) that the pump is much stronger than the signal, we take $E_{0x,2}$ to be constant and the phase-matching to be perfect, i.e., $n_1 = n_2$; the solution of Eq.(28) may then be written as

$$E_{0x,1}(z) = E_{0x,1}(0)\cosh(\zeta z) + C\sinh(\zeta z), \quad \text{where} \quad \zeta = \frac{\omega_1 \chi^{(2)}|E_{0x,2}|}{2n_1 c}. \tag{32a}$$

Substituting Eq.(32a) into Eq.(28) yields the remaining unspecified coefficient $C$, as follows:

$$C = iE^*_{0x,1}(0)E_{0x,2}/|E_{0x,2}|. \tag{32b}$$

Observe that the character of the signal amplitude $E_{0x,1}(z)$ given by Eq.(32) differs substantively from that obtained in the case of non-degenerate parametric amplification and given by Eq.(21). The phase difference between the initial signal and the pump plays a pivotal role in the present (degenerate) case, as is readily inferred from Eq.(32). The signal is now a mixture of $e^{\zeta z}$ and $e^{-\zeta z}$ functions, which, depending on the phase difference between the initial signal and pump amplitudes, $E_{0x,1}(0)$ and $E_{0x,2}(0)$, could exhibit amplification or attenuation. Whereas amplification involves a transfer of electromagnetic energy from the pump to the signal, in the case of degenerate parametric attenuation (or de-amplification) it is the signal that pours its energy into the pump. In either case, the total optical energy is conserved since $\chi^{(1)}$ and $\chi^{(2)}$ have been assumed at the outset to be real-valued.

Writing $E_{0x,2} = |E_{0x,2}|e^{i\varphi_2}$, the complete expression of the emergent $E$-field at the exit facet of a degenerate parametric amplifier of length $L$ is obtained from Eq.(32), as follows:

$$\boldsymbol{E}_1(x,y,z=L,t) = [\cosh(\zeta L)\,E_{0x,1}(0) + ie^{i\varphi_2}\sinh(\zeta L)\,E^*_{0x,1}(0)]\hat{\boldsymbol{x}}\,e^{i[n_1\omega_1(L/c)-\omega_1 t]}. \tag{33}$$

This purely classical expression has the correct structure to be turned into the corresponding quantum operator for the $E$-field emerging at the output of the amplifier, namely,

$$\widehat{\boldsymbol{E}}^{(+)}_1(L) = e^{in_1\omega_1 L/c}\bigl[\cosh(\zeta L)\,\hat{E}^{(+)}_{x,1}(0) + ie^{i\varphi_2}\sinh(\zeta L)\,\hat{E}^{(-)}_{x,1}(0)\bigr]\hat{\boldsymbol{x}}\,. \tag{34}$$

It is easy to verify that this single-mode $E$-field operator at the amplifier's output has the correct (i.e., canonical) commutation relations given an input beam in the same single-mode; in particular, considering that $\cosh^2(\zeta L) - \sinh^2(\zeta L) = 1$, we find

$$\bigl[\hat{E}^{(+)}_{x,1}(L), \hat{E}^{(-)}_{x,1}(L)\bigr] = \bigl[\hat{E}^{(+)}_{x,1}(0), \hat{E}^{(-)}_{x,1}(0)\bigr]. \tag{35}$$

The situation encountered in the present example is exceptional, in that it permits noiseless amplification of the input signal. Nevertheless, it does not violate the general principle that quantum amplification must always be accompanied by additional quantum noise, the reason being that the input beam in the present example is required to have a well-defined phase (relative to the pump beam) for the gain to materialize.[8]

**6. Quantum noise accompanying attenuation and amplification of light beams**. In quantum optics, one can represent the attenuation of an incoming EM wave by means of a lossless and symmetric beam-splitter having Fresnel reflection and transmission coefficients $\rho$ and $\tau$, with $|\rho|^2 + |\tau|^2 = 1$ and $\varphi_\rho = \varphi_\tau + 90°$; see Fig.2.[6-8,12] Thus, the annihilation operator at the exit port 3 of the splitter relates to the corresponding operators at the entrance ports 1 and 2 via the linear combination $\hat{a}_3 = \tau \hat{a}_1 + \rho \hat{a}_2$. Similarly, the annihilation operator at the exit port 4 relates to $\hat{a}_1$ and $\hat{a}_2$ via $\hat{a}_4 = \rho \hat{a}_1 + \tau \hat{a}_2$, although, for purposes of the present discussion, the emergent beam at the exit port 4 is irrelevant and shall be ignored.

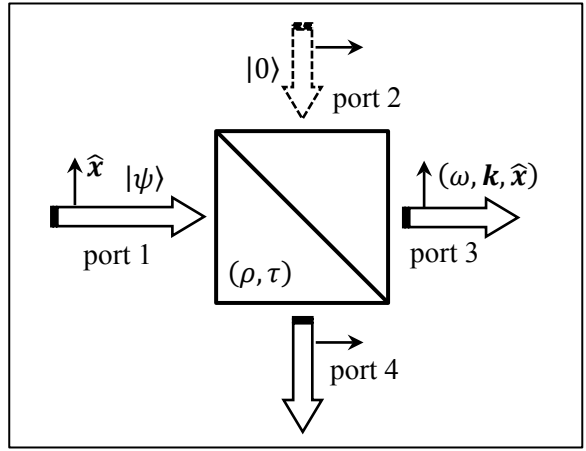

**Fig.2**. A propagating plane-wave mode $(\omega, \boldsymbol{k}, \hat{\boldsymbol{x}})$ in the state $|\psi\rangle$ enters through port 1 of a lossless and symmetric beam-splitter, whose Fresnel reflection and transmission coefficients are specified as $\rho$ and $\tau$, respectively. An attenuated beam exits through port 3, with added noise due to quantum fluctuations that come through the vacuum port 2. The positive-frequency $E$-field operator at the exit port 3 is given by Eq.(36), where the annihilation operator $\hat{a}_3$ is the linear combination $\tau \hat{a}_1 + \rho \hat{a}_2$ of the corresponding operators at the input ports 1 and 2.

We assume that the incident beam entering through port 1 is the single-mode propagating plane-wave $(\omega, \boldsymbol{k}, \hat{\boldsymbol{e}})$ in the pure state $|\psi\rangle$, and that nothing enters through port 2 except, of course, the ever-present vacuum state $|0\rangle$. To simplify the notation, we take the field's polarization to be linear along the $x$-axis (i.e., $\hat{\boldsymbol{e}} = \hat{\boldsymbol{x}}$) and proceed to write the emergent positive-frequency $E$-field operator in port 3 as follows:

$$\hat{E}_3^{(+)}(\boldsymbol{r}, t) = \mathrm{i} \left( \frac{\hbar \omega}{2 \varepsilon_0 V} \right)^{½} \hat{\boldsymbol{x}} e^{\mathrm{i}(\boldsymbol{k} \cdot \boldsymbol{r} - \omega t)} \hat{a}_3. \qquad (36)$$

The commutator of $\hat{E}_{3,x}^{(+)}$ with its conjugate transpose (or adjoint) $\hat{E}_{3,x}^{(-)}$ is readily found to be

$$[\hat{E}_{3,x}^{(+)}, \hat{E}_{3,x}^{(-)}] = \left( \frac{\hbar \omega}{2 \varepsilon_0 V} \right) [\hat{a}_3, \hat{a}_3^\dagger] = \left( \frac{\hbar \omega}{2 \varepsilon_0 V} \right) [(\tau \hat{a}_1 + \rho \hat{a}_2), (\tau^* \hat{a}_1^\dagger + \rho^* \hat{a}_2^\dagger)]$$

$$= \left( \frac{\hbar \omega}{2 \varepsilon_0 V} \right) (|\tau|^2 [\hat{a}_1, \hat{a}_1^\dagger] + |\rho|^2 [\hat{a}_2, \hat{a}_2^\dagger]) = \frac{\hbar \omega}{2 \varepsilon_0 V}. \qquad (37)$$

This is the correct (i.e., canonical) commutation relation for the attenuated $E$-field of the beam that arrives at port 1 and proceeds to emerge from port 3 of the beam-splitter. The positive-frequency $E$-field operator $\hat{\boldsymbol{E}}_3^{(+)}$ of Eq.(36) with $\hat{a}_3 = \tau \hat{a}_1 + \rho \hat{a}_2$, together with the field's conjugate transpose operator $\widehat{E}_3^{(-)}$, act on the composite input state $|\psi\rangle_1|0\rangle_2$ to reveal certain properties of the attenuated beam. Our notation will be a little simplified if we take the attenuation coefficient $\tau$ to be real-valued, in which case the (purely imaginary) reflection coefficient $\rho$ can be written as $\sqrt{\tau^2-1}$.

Placing a photodetector at the splitter's exit port 3 and proceeding to compute the expected value of the intensity of the attenuated wave, we find

$$\bar{I}_3 = \langle 0_2, \psi_1|\widehat{E}_{3,x}^{(-)}\widehat{E}_{3,x}^{(+)}|\psi_1, 0_2\rangle$$
$$= \left(\frac{\hbar\omega}{2\varepsilon_0 V}\right)\langle 0_2, \psi_1|\tau^2\hat{a}_1^\dagger\hat{a}_1 + i\tau\sqrt{1-\tau^2}(\hat{a}_1^\dagger\hat{a}_2 - \hat{a}_2^\dagger\hat{a}_1) + (1-\tau^2)\hat{a}_2^\dagger\hat{a}_2|\psi_1, 0_2\rangle$$
$$= \tau^2\left(\frac{\hbar\omega}{2\varepsilon_0 V}\right)\langle\psi|\hat{a}_1^\dagger\hat{a}_1|\psi\rangle = \tau^2\bar{I}_1. \tag{38}$$

As for the expected value of the squared intensity, application of standard operator algebra yields

$$\overline{I_3^2} = \langle 0_2, \psi_1|\widehat{E}_{3,x}^{(-)}\widehat{E}_{3,x}^{(+)}\widehat{E}_{3,x}^{(-)}\widehat{E}_{3,x}^{(+)}|\psi_1, 0_2\rangle = \left(\frac{\hbar\omega}{2\varepsilon_0 V}\right)^2\tau^4\langle\psi|\hat{a}_1^\dagger\hat{a}_1^\dagger\hat{a}_1\hat{a}_1|\psi\rangle + \left(\frac{\hbar\omega}{2\varepsilon_0 V}\right)\tau^2\bar{I}_1. \tag{39}$$

In the case of a quasi-classical (or Glauber) coherent state entering through port 1, namely, $|\psi\rangle_1 = |\gamma\rangle_1$, we will have $\bar{I}_1 = (\hbar\omega/2\varepsilon_0 V)|\gamma|^2$ and $\langle\gamma|\hat{a}_1^\dagger\hat{a}_1^\dagger\hat{a}_1\hat{a}_1|\gamma\rangle = |\gamma|^4$, in which case the variance of the intensity at the exit port 3 will be $(\hbar\omega/2\varepsilon_0 V)\tau^2\bar{I}_1$, in agreement with what one expects from an emergent coherent beam $|\tau\gamma\rangle_3$. In contrast, a noiseless number-state $|n\rangle$ arriving at port 1 emerges at port 3 with an average intensity $\bar{I}_3 = \tau^2\bar{I}_1 = (n\hbar\omega/2\varepsilon_0 V)\tau^2$, accompanied by a substantial noise component, namely, $\text{var}(I_3) = (\sqrt{n}\hbar\omega/2\varepsilon_0 V)^2\tau^2(1-\tau^2)$.

If a beam is amplified (as opposed to being attenuated) with a gain coefficient of $g$ (assumed here to be real and greater than 1), its $E$-field's canonical commutation relation can be restored by augmenting the amplified beam's positive frequency $E$-field operator with a vacuum contribution, as follows:[8,13]

$$\widehat{E}_3^{(+)}(\boldsymbol{r}, t) = i\left(\frac{\hbar\omega}{2\varepsilon_0 V}\right)^{1/2}\hat{\boldsymbol{e}}\, e^{i(\boldsymbol{k}\cdot\boldsymbol{r}-\omega t)}\left(g\hat{a}_1 + \sqrt{g^2-1}\,\hat{a}_2^\dagger\right). \tag{40}$$

The assumed model here is that the amplifier has an entrance port 1 and a vacuum port 2, through which enters the composite single-mode beam in the state $|\psi\rangle_1|0\rangle_2$. Note that the annihilation operator $\hat{a}_3$ corresponding to exit port 3 contains an anticipated term $g\hat{a}_1$ (associated with a linear gain coefficient $g$) plus an additional (noise) term $\sqrt{g^2-1}\,\hat{a}_2^\dagger$ contributed by the vacuum port 2. The similarity of the $E$-field operator appearing in Eq.(40) to those of parametric amplification given by Eqs.(25) is not coincidental. The commutator of $\widehat{E}_3^{(+)}$ of Eq.(40) and its conjugate transpose (or adjoint) $\widehat{E}_3^{(-)}$ is readily found to be

$$\left[\widehat{E}_3^{(+)}, \widehat{E}_3^{(-)}\right] = \left(\frac{\hbar\omega}{2\varepsilon_0 V}\right)[\hat{a}_3, \hat{a}_3^\dagger] = \left(\frac{\hbar\omega}{2\varepsilon_0 V}\right)\left(g^2[\hat{a}_1, \hat{a}_1^\dagger] - (g^2-1)[\hat{a}_2, \hat{a}_2^\dagger]\right) = \frac{\hbar\omega}{2\varepsilon_0 V}. \tag{41}$$

If we now imagine a photodetector placed at the exit port 3 and proceed to compute the expected value of the intensity of the amplified wave, we find

$$\bar{I}_3 = \langle 0_2, \psi_1|\widehat{\boldsymbol{E}}_3^{(-)}\cdot\widehat{\boldsymbol{E}}_3^{(+)}|\psi_1, 0_2\rangle$$
$$= \left(\frac{\hbar\omega}{2\varepsilon_0 V}\right)\langle 0_2, \psi_1|g^2\hat{a}_1^\dagger\hat{a}_1 + g\sqrt{g^2-1}(\hat{a}_1^\dagger\hat{a}_2^\dagger + \hat{a}_2\hat{a}_1) + (g^2-1)\hat{a}_2\hat{a}_2^\dagger|\psi_1, 0_2\rangle$$
$$= g^2\bar{I}_1 + \left(\frac{\hbar\omega}{2\varepsilon_0 V}\right)(g^2-1). \tag{42}$$

Thus, aside from the anticipated $g^2 \bar{I}_1$, the expected value (or average) of the amplified intensity contains an additional term arising from vacuum fluctuations in port 2. As for the expected value of the squared intensity, standard operator algebra yields

$$\overline{I_3^2} = \langle 0_2, \psi_1|(\hat{\boldsymbol{E}}_3^{(-)} \cdot \hat{\boldsymbol{E}}_3^{(+)})(\hat{\boldsymbol{E}}_3^{(-)} \cdot \hat{\boldsymbol{E}}_3^{(+)})|\psi_1, 0_2\rangle$$
$$= \left(\frac{\hbar\omega}{2\varepsilon_0 V}\right)^2 \langle\psi|g^4(\hat{a}_1^\dagger \hat{a}_1^\dagger \hat{a}_1 \hat{a}_1) + (4g^4 - 3g^2)(\hat{a}_1^\dagger \hat{a}_1) + (2g^2 - 1)(g^2 - 1)|\psi\rangle. \qquad (43)$$

The variance of the intensity at port 3 is now found from Eq.(43) and the first term on the right-hand side of Eq.(42), as follows:

$$\text{var}(I_3) = \overline{I_3^2} - (g^2 \bar{I}_1)^2. \qquad (44)$$

In the special case of a quasi-classical (or Glauber) coherent beam entering through port 1 of an amplifier, we have $|\psi\rangle_1 = |\gamma\rangle_1$ and, therefore,

$$\bar{I}_1 = \left(\frac{\hbar\omega}{2\varepsilon_0 V}\right)|\gamma|^2, \qquad (45)$$

$$\text{var}(I_3) = \left(\frac{\hbar\omega}{2\varepsilon_0 V}\right)^2 [(4g^4 - 3g^2)|\gamma|^2 + (2g^4 - 3g^2 + 1)]. \qquad (46)$$

For the amplified coherent beam emerging at the exit port 3, the ratio of the signal $g^2 \bar{I}_1$ to the standard deviation of $I_3$ (i.e., the noise) will be $|\gamma|$ if $g = 1$, but it approaches $|\gamma|/2$ for $g \gg 1$ provided that $|\gamma|$ is sufficiently large. This 3 dB drop in the signal-to-noise ratio of a coherent beam in consequence of its amplification is an inescapable result of vacuum fluctuations that appear as a fundamental source of quantum noise in optical amplifiers.

**7. Parametric fluorescence**. This is a purely quantum-mechanical phenomenon in which an intense laser beam passing through a crystal such as KDP (i.e., potassium dihydrogen phosphate $KH_2PO_4$, a $\chi^{(2)}$ material) creates a pair of photons whose energies $\hbar\omega_1$ and $\hbar\omega_2$ add up to the energy $\hbar\omega_3$ of a single photon of the incident laser beam; see Fig.3. It is possible for the photons in a pair to have identical energies, in which case $\omega_1 = \omega_2 = \frac{1}{2}\omega_3$. The propagation directions of the emergent twin photons along $\boldsymbol{k}_1$ and $\boldsymbol{k}_2$ are determined by the phase-matching condition inside the crystal. While the incident $(\omega_3, \boldsymbol{k}_3)$ photons are responsible for the excitation of the host medium, the cascade emission of the $(\omega_1, \boldsymbol{k}_1)$ and $(\omega_2, \boldsymbol{k}_2)$ photon pair has no classical analog, being reminiscent of (and akin to) the spontaneous emission process from excited atoms.

In a typical experiment,[14] one selects the conjugate directions of the twin-photon pair by placing small-aperture diaphragms in the path of individual photons. The range of frequencies of each emitted photon is also restricted by band-pass filters, one of which transmits a narrow band of frequencies centered at $\omega_1$ (and propagating along $\boldsymbol{k}_1$) to detector $D_1$, while the other one allows a narrow band centered at $\omega_2$ (and propagating along $\boldsymbol{k}_2$) to pass through to detector $D_2$.

The twin-photon pairs typically have a range of frequencies $(\omega_1 - \delta, \omega_2 + \delta)$, where $\delta$ has a narrow distribution (such as one with a slim bell-shaped profile) centered at zero. The joint state of any given photon pair may thus be expressed by the following superposition:

$$|\psi\rangle = \sum_\ell c(\delta) |1_{(\boldsymbol{k}_1, \omega_1 - \delta)}, 1_{(\boldsymbol{k}_2, \omega_2 + \delta)}\rangle. \qquad (47)$$

Here, the index $\ell$ enumerates the pair of electromagnetic field modes associated with each value of $\delta$, while $c(\delta)$ is the probability amplitude for each photon pair to be produced in a given photoemission event. Assuming the two photons of a pair follow essentially equivalent paths to the detectors, the probability amplitude that photon 1 is captured by detector $D_1$ at time $t_1$ is proportional to $e^{-i(\omega_1-\delta)t_1}$, while that of photon 2 being captured by $D_2$ at time $t_2$ is proportional to $e^{-i(\omega_2+\delta)t_2}$. The amplitude for detecting the first photon of a pair at $t_1$ and the second one at $t_2$ is thus found to be proportional to

$$e^{-i(\omega_1 t_1+\omega_2 t_2)}\sum_\ell c(\delta)e^{i\delta(t_1-t_2)}. \quad \leftarrow \boxed{\text{Note: exponent is } not \text{ a } \delta\text{-function}} \quad (48)$$

This summing of the probability amplitudes presumes that the various joint detection events at fixed instants of time $(t_1, t_2)$ are indistinguishable from one another, despite the fact that each value of $\delta$ corresponds to distinct frequencies of the photons that arrive at each detector. The fact remains, however, that the specificity of the photon arrival times is incompatible with the knowledge of the corresponding photon energies, i.e., $\hbar(\omega_1-\delta)$ and $\hbar(\omega_2+\delta)$, or, equivalently, the knowledge of their frequencies.

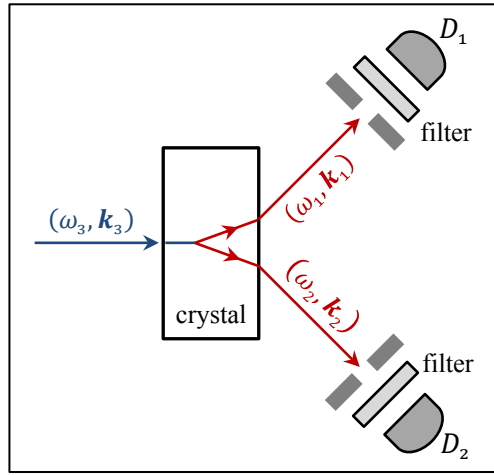

**Fig.3**. An intense light beam of frequency $\omega_3$, say, from an argon ion laser, propagates along its $k$-vector, $\boldsymbol{k}_3$, enters a KDP crystal, and produces the twin-photon pair $(\omega_1, \boldsymbol{k}_1)$ and $(\omega_2, \boldsymbol{k}_2)$, with $\omega_1+\omega_2=\omega_3$. The photon pairs thus produced by parametric fluorescence and selected for propagation along conjugate directions (by means of small-aperture diaphragms placed in the path of individual photons), pass through narrowband filters and are subsequently captured by photodetectors $D_1$ and $D_2$.[14]

The sum over $\ell$ in Eq.(48) may be treated as an integral over $\delta$, resulting in the Fourier transform $\tilde{c}(t_1-t_2)$ of $c(\delta)$. Considering that $c(\delta)$ is typically a bell-shaped function of width $\sim 10^{13}$ Hz, we conclude that $\tilde{c}(t_1-t_2)$ is similarly bell-shaped, albeit with a width of $\sim 10^{-13}$ sec. The process of parametric fluorescence is thus seen to create twin-photon pairs comprised of two nearly simultaneously created photons (e.g., $|t_1-t_2|\sim 0.1$ ps).

A single photodetection event by detector $D_1$ at time $t_1$ will have amplitude $c(\delta)e^{-i(\omega_1-\delta)t_1}$ for any given value of $\delta$. Such events, however, are (in principle) distinguishable, since the pair's second photon can be analyzed for its frequency, $\omega_2+\delta$, thus yielding the precise value of $\delta$. One must, therefore, add up the corresponding probabilities $|c(\delta)e^{-i(\omega_1-\delta)t_1}|^2=|c(\delta)|^2$ of single-photon detection events by $D_1$ at $t=t_1$ to arrive at $\int|c(\delta)|^2\mathrm{d}\delta$, which is clearly a constant, i.e., independent of time. The same argument, of course, applies to single-photon detection events by $D_2$ at $t=t_2$. Thus, each detector may report the capture of a photon at any time while the exciting light source (frequency $=\omega_3$) is turned on. However, the emission of the photon captured by $D_1$ should be coincident (or nearly so) with the emission and subsequent capture of its twin by $D_2$.

An alternative way of arriving at the above conclusions entails the application of operators for single-photon detection at $D_1$ or $D_2$, and also joint photon-pair detection at both detectors. The positive-frequency $E$-field operator for $D_1$ is

$$\hat{E}_1^{(+)}(\boldsymbol{r},t) = \sum_\ell \mathrm{i}(\hbar\omega_{\ell,1}/2\varepsilon_0 V)^{\frac{1}{2}} e^{\mathrm{i}(\boldsymbol{k}_{\ell,1}\cdot\boldsymbol{r}-\omega_{\ell,1}t)}\hat{a}_{\ell,1}. \tag{49}$$

The corresponding operator for $D_2$ is similar, of course, with the subscripts 1 replaced by 2. Invoking standard operator algebra, the detection rate at $D_1$ is proportional to

$$\langle\psi|\hat{E}_1^{(-)}(\boldsymbol{r}_1,t_1)\hat{E}_1^{(+)}(\boldsymbol{r}_1,t_1)|\psi\rangle = \|\hat{E}_1^{(+)}(\boldsymbol{r}_1,t_1)|\psi\rangle\|^2. \tag{50}$$

The operator $\hat{E}_1^{(+)}$ of Eq.(49) acting on the state $|\psi\rangle$ of Eq.(47) yields

$$\hat{E}_1^{(+)}(\boldsymbol{r}_1,t_1)|\psi\rangle = \mathrm{i}(\hbar/2\varepsilon_0 V)^{\frac{1}{2}} e^{\mathrm{i}(\boldsymbol{k}_1\cdot\boldsymbol{r}_1-\omega_1 t_1)} \sum_\ell (\omega_1-\delta)^{\frac{1}{2}} e^{\mathrm{i}\delta t_1} c(\delta)|0, 1_{(\boldsymbol{k}_2,\omega_2+\delta)}\rangle. \tag{51}$$

Considering that $\langle 0,1_{(\boldsymbol{k}_2,\omega_2+\delta')}|0,1_{(\boldsymbol{k}_2,\omega_2+\delta)}\rangle$ equals zero when $\delta \neq \delta'$ and 1.0 when $\delta = \delta'$, Eq.(50) in conjunction with Eq.(51) now reveals the detection rate at $D_1$ to be proportional to

$$(\hbar/2\varepsilon_0 V)\int(\omega_1-\delta)|c(\delta)|^2 \mathrm{d}\delta \cong (\hbar\omega_1/2\varepsilon_0 V)\int|c(\delta)|^2 \mathrm{d}\delta. \tag{52}$$

As for the joint detection rate at $D_1$ and $D_2$, we write

$$\langle\psi|\hat{E}_1^{(-)}(\boldsymbol{r}_1,t_1)\hat{E}_2^{(-)}(\boldsymbol{r}_2,t_2)\hat{E}_2^{(+)}(\boldsymbol{r}_2,t_2)\hat{E}_1^{(+)}(\boldsymbol{r}_1,t_1)|\psi\rangle = \|\hat{E}_2^{(+)}(\boldsymbol{r}_2,t_2)\hat{E}_1^{(+)}(\boldsymbol{r}_1,t_1)|\psi\rangle\|^2. \tag{53}$$

The action of $\hat{E}_2^{(+)}(\boldsymbol{r}_2,t_2)\hat{E}_1^{(+)}(\boldsymbol{r}_1,t_1)$ on the state $|\psi\rangle$ of Eq.(47) yields

$$\hat{E}_2^{(+)}(\boldsymbol{r}_2,t_2)\hat{E}_1^{(+)}(\boldsymbol{r}_1,t_1)|\psi\rangle$$
$$= -(\hbar/2\varepsilon_0 V) e^{\mathrm{i}(\boldsymbol{k}_1\cdot\boldsymbol{r}_1+\boldsymbol{k}_2\cdot\boldsymbol{r}_2-\omega_1 t_1-\omega_2 t_2)} \sum_\ell (\omega_1-\delta)^{\frac{1}{2}}(\omega_2+\delta)^{\frac{1}{2}} e^{\mathrm{i}\delta(t_1-t_2)} c(\delta)|0,0\rangle$$
$$\cong -(\hbar\sqrt{\omega_1\omega_2}/2\varepsilon_0 V) e^{\mathrm{i}(\boldsymbol{k}_1\cdot\boldsymbol{r}_1+\boldsymbol{k}_2\cdot\boldsymbol{r}_2-\omega_1 t_1-\omega_2 t_2)} \sum_\ell e^{\mathrm{i}\delta(t_1-t_2)} c(\delta)|0,0\rangle. \tag{54}$$

The joint detection rate at $D_1$ and $D_2$ is now obtained from Eq.(53) in conjunction with Eq.(54) as

$$\|\hat{E}_2^{(+)}(\boldsymbol{r}_2,t_2)\hat{E}_1^{(+)}(\boldsymbol{r}_1,t_1)|\psi\rangle\|^2 \cong (\hbar\sqrt{\omega_1\omega_2}/2\varepsilon_0 V)^2 \left|\int c(\delta) e^{\mathrm{i}\delta(t_1-t_2)} \mathrm{d}\delta\right|^2. \tag{55}$$

Once again, it is seen that the joint probability of detecting the twin-photon pair at $(t_1, t_2)$ is proportional to $|\tilde{c}(t_1-t_2)|^2$, where $\tilde{c}(t)$ is the Fourier transform of $c(\delta)$.

**8. Concluding remarks**. This paper has described some of the fundamental, albeit elementary, aspects of nonlinear optical phenomena, with an eye toward facilitating the pedagogical transition from a classical to a quantum-optical understanding of the underlying physical concepts and methods. A most interesting application of nonlinear optics in recent years has been the injection of squeezed vacuum radiation (created by degenerate optical-parametric amplification, also known as optical-parametric down-conversion) into the *Laser Interferometer Gravitational-Wave Observatory* (LIGO).[15,16] This has resulted in a substantial improvement of the signal-to-noise ratio of the photo-detectors, thus enhancing the overall sensitivity of the system and expanding the detection range of gravitational-wave observations.

As more powerful lasers become available and the demand for high-precision metrology, intricate optical micro-manipulation, accurate optical trapping, etc., continues to grow, the need for novel nonlinear optical tools and innovative techniques is bound to surpass the most realistic expectations. There exists a vast literature covering the science and the technological applications of nonlinear optics, to which the interested reader is referred for further information.[1-8,17-20]